\documentclass[12pt]{article}
\usepackage{times,epsfig,cite,amssymb,amsmath}

\begin{document}

\title{Coherent states approach to Penning trap}

\author{David J. Fern\'andez C. and Mercedes Vel\'azquez \\
Departamento de F\'{\i}sica, Cinvestav \\ A.P. 14-740, 07000
M\'exico D.F., Mexico}

\date{}

\maketitle

\begin{abstract}
By using a matrix technique, which allows to identify directly the
ladder operators, the Penning trap coherent states are derived as
eigenstates of the appropriate annihilation operators. These states
are compared with the ones obtained through the displacement
operator. The associated wave functions and mean values for some
relevant operators in these states are also evaluated. It turns out
that the Penning trap coherent states minimize the Heisenberg
uncertainty relation.
\end{abstract}


\section{Introduction}

The coherent states (CS) approach to quantum physical systems
\cite{ks85,pe86,gaa00} constitutes nowadays an alternative to the
standard method, which address the same problem in terms of energy
eigenstates and eigenvalues. Along the years the CS have been
derived for plenty of Hamiltonians having either a ground or a top
state, and some of them admit a group theoretical construction in
which this state is acted on by an appropriate displacement operator
\cite{pe86,rr00}. However, there exist interesting physical systems
for which the Hamiltonians have neither ground nor top state
\cite{mf89,cc05}, but it is required anyway a systematic technique
to build up the corresponding CS. One of those systems consists of a
charged particle in an ideal Penning trap \cite{bg86,fe92}. Such an
arrangement, sometimes called Geonium atom, has been largely used to
perform high precision measurements of fundamental properties of
particles \cite{bg86}. Moreover, it could be used to test and/or
control some intrinsically quantum phenomena as entanglement,
decoherence, wavepacket reduction, etc \cite{fe92,fm94,ma97}.

In this paper we are going to address, from a coherent states
viewpoint, the quantum motion of a charged particle in a Penning
trap. With this aim, in section 2 we will present some generalities
of the standard coherent states. In sections 3 and 4 we will
introduce the Penning trap Hamiltonian and discuss its corresponding
algebraic structure. It will be shown that the system possesses a
certain ``extremal'' state, which plays the role of a ground state
although there is not a minimum energy eigenvalue. In section 5 we
will construct the wavefunction associated to the extremal state,
while in section 6 we will perform the corresponding CS
construction. The mean values of some physical quantities in the CS
will be calculated in section 7. Finally, in section 8 our
conclusions will be presented.

\section{Standard coherent states}

Glauber definitions of CS are based on properties of the harmonic
oscillator \cite{gl63a}, which have been applied to several
different systems (see e.g. \cite{ks85,pe86,gaa00}):

\noindent (1) The CS $\vert z\rangle$ are eigenstates of the
annihilation operator $a$:
\begin{eqnarray}
& a \vert z\rangle = z\vert z\rangle, \quad z \in {\mathbb C}.
\label{eigena}
\end{eqnarray}

\noindent (2) They arise from acting the displacement operator on
the ground state,
\begin{eqnarray}
& \vert z\rangle = D(z)\vert\psi_0\rangle, \quad D(z) =
\exp(za^\dagger - z^* a), \label{gsdisplaced}
\end{eqnarray}
$a^\dagger$ being the creation operator.

\noindent (3) The CS satisfy the minimum Heisenberg uncertainty
relation for $X$ and $P$,
\begin{eqnarray}
& (\Delta X)_z(\Delta P)_z = \hbar/2 , \label{dxdpscs}
\end{eqnarray}
where $(\Delta{\mathcal O})_z^2 = \langle z \vert ({\mathcal O} -
\langle{\mathcal O}\rangle_z)^2 \vert z \rangle = \langle{\mathcal
O}^2\rangle_z - \langle{\mathcal O}\rangle_z^2$ is the mean square
deviation of an observable ${\mathcal O}$ in the state $\vert
z\rangle$.

It is worth to notice an additional property of the standard CS,
which is relevant since some authors consider it as the fourth CS
definition. It is the completeness relationship $\frac{1}{\pi} \int
\vert z\rangle \langle z \vert d^2 z = {\bf 1}$, where ${\bf 1}$ is
the identity operator. In fact, the CS are overcomplete in the sense
that for any convergent sequence of complex numbers $z_n$ the
corresponding CS $\vert z_n\rangle$ form a complete set
\cite{bbgk71}.

For systems different from the harmonic oscillator, these
definitions lead to different sets of CS. In this paper we will use
the first and second definitions to find the CS for a charged
particle in an ideal Penning trap; we will show that they satisfy as
well equation (\ref{dxdpscs}).

\section{Penning trap Hamiltonian and the matrix $\pmb{\Lambda}$}

Let us consider a spinless particle of mass $m$ and electric charge
$e$ inside of an ideal Penning trap, i.e., under the influence of a
constant homogeneous magnetic field pointing along the $z$-direction
$\vec{B}=B\hat{k}$, and a static electric field
$\vec{E}=-\vec{\nabla} \Phi(\vec{r})$, both arising from the
following vector and quadrupole scalar potentials:
\begin{eqnarray}
\vec{A}(\vec{r}) = - \frac12 \vec{r}\times\vec{B}, \qquad
\Phi(\vec{r})=\frac{\Phi_0}{d^2}(x^2+y^2-2z^2).
\end{eqnarray}
Throughout this paper, the small letters
$\vec{r},\vec{p},x,y,z,p_x,p_y,p_x$ will denote either classical
coordinates and momenta or the eigenvalues associated to the
corresponding quantum operators, the last ones being represented by
capital letters $\vec{R},\vec{P},X,Y,Z,P_x,P_y,P_z$. The Hamiltonian
describing our quantum system is given by
\begin{eqnarray}
\hskip-1.6cm H & \! = \! & \frac{1}{2m} \big(
\vec{P}-\frac{e}{c}\vec{A}(\vec{R})\big)^2 + e \Phi(\vec{R}) \! = \!
\frac{\vec{P}^2}{2m} + b L_z + \frac{m}2\left[(b^2 + v)(X^2+Y^2) - 2
v Z^2 \right],
\end{eqnarray}
where $\vec{L} = \vec{R}\times \vec{P} $ is the angular momentum
operator, $b=-\frac{eB}{2mc}$, $v=\frac{2e\Phi_0}{md^2}$ and we take
by simplicity $b>0$. To ensure that the particle is trapped inside
the cavity, some restrictions on the parameters $b,\ v$ have to be
taken: first of all $v<0$ in order that the $z$-motion is bounded
(so that this mode is characterized by a standard oscillator
Hamiltonian). However, the corresponding repulsive oscillators in
the $x-y$ plane do not have to destroy the trapped motion induced by
the magnetic field, which is achieved by taking $b^2+v>0$.

From now on we will assume that $m=1$ and $\hbar = 1$. Note that
this assumption is equivalent to the following procedure: (i) first
making the operator changes $\hat{R_i} = {R_i} \sqrt{m/\hbar}$,
$\hat{P_i} = {P_i}/ \sqrt{m\hbar}, \ i=1,2,3$, $\hat H = H/\hbar$;
(ii) then dropping the hats in order to simplify the notation. Thus,
the Hamiltonian we are dealing with reads
\begin{eqnarray}
H & = & \frac{\vec{P}^2}{2} + b L_z + \frac{1}2\left[(b^2 +
v)(X^2+Y^2) - 2 v Z^2 \right],
\end{eqnarray}
where $[R_i,P_j] = i \delta_{ij}$.

It is useful to work in the Heisenberg picture in which the
evolution of the operator vector $\eta(t) = U^\dagger (t) \eta U(t)$
is simply determined from a matrix equation:
\begin{equation}\label{evolu temp}
\frac{d \eta(t)}{dt} = U^\dagger(t)[iH, \eta]U(t) =
U^\dagger(t)\pmb{\Lambda} \eta U(t) =\pmb{\Lambda}\eta(t) \quad
\Rightarrow \quad \eta(t) = e^{\pmb{\Lambda}t}\eta ,
\end{equation}
where $\eta = (\vec{R}, \ \vec{P})^{\rm T}$ involves the observables
$\vec{R}, \ \vec{P}$ in the Schr\"odinger picture, the superindex
${}^{\rm T}$ denotes to transpose the involved vector, $U(t)$ is the
evolution operator such that $U(0) = {\bf 1}$. The calculation of
$[iH, \eta] = \pmb{\Lambda}\eta$ leads to
\begin{eqnarray}\label{Lambda tp}
& \pmb{\Lambda}=\left( \begin{array}{cccccc}
                0&-b&0&1&0&0\\
                b&0&0&0&1&0\\
                0&0&0&0&0&1\\
                -b^2-v&0&0&0&-b&0\\
                0&-b^2-v&0&b&0&0\\
                0&0&2v&0&0&0
                \end{array} \right).
\end{eqnarray}
Let us find next the right ($u$) and left ($f$) eigenvectors of the
matrix $\pmb{\Lambda}$, which are called eigenvectors and eigenforms
respectively. Since $\pmb{\Lambda}$ is non Hermitian, the eigenforms
$f$ are not necessarily the adjoints of the eigenvectors $u$. In
order to determine both, we solve in the first place the
characteristic equation of $\pmb{\Lambda}$:
\begin{eqnarray}\label{ecuacion caracteristica}
& \vert \pmb{\Lambda} - \lambda {\bf 1}\vert =
\lambda^6+4b^2\lambda^4-v(8b^2+3v)\lambda^2-2v^3=0.
\end{eqnarray}
Thus, the eigenvalues are $\pm\lambda_1 = \pm i\omega_1,\pm\lambda_2
= \pm i\omega_2,\pm\lambda_3 = \pm i\omega_3$, where
\begin{eqnarray}\label{eigenvalores}
\hskip-0.8cm
    \omega_1=b+\sqrt{b^2+v}, \quad
    \omega_2 = b-\sqrt{b^2+v}, \quad \omega_3&=&\sqrt{-2v}.
\end{eqnarray}
We label as $u_k$, $u_k^*$ and $f_k$, $f_k^*$ the eigenvectors and
eigenforms associated to the eigenvalues $\lambda_k$, $\lambda_k^* =
- \lambda_k$ respectively, i.e., $\pmb{\Lambda} u_k = \lambda_k
u_k$, $\pmb{\Lambda} u_k^* = - \lambda_k u_k^*$, $f_k \pmb{\Lambda}
= \lambda_k f_k$, $f_k^* \pmb{\Lambda} = -\lambda_k f_k^*$,
$k=1,2,3$, the $^*$ denoting complex conjugation. An explicit
calculation leads to:
\begin{eqnarray}
    &  u_1  =  s_1
        \left(\frac{1}{\sqrt{b^2+v}},\frac{-i}{\sqrt{b^2+v}},0,i,1,0\right)\!^{\rm
        T}, & \hskip0.5cm
f_1 = t_1\left(\sqrt{b^2+v},
    i\sqrt{b^2+v},0,-i,1,0 \right), \nonumber \\
    &  u_2  =  s_2
        \left(\frac{-1}{\sqrt{b^2+v}},\frac{i}{\sqrt{b^2+v}},0,i,1,0\right)\!^{\rm T} \! ,
        & \hskip0.5cm f_2 = t_2\left(-\sqrt{b^2+v},
    -i\sqrt{b^2+v},0,-i,1,0 \right), \nonumber \\
    &  u_3  =  s_3
        \left(0,0, \frac{-i}{\sqrt{-2v}},0,0,1\right)\!^{\rm T}, &
    \hskip0.5cm f_3 = t_3\left(0,0,i\sqrt{-2v},0,0,1 \right) \nonumber,
\end{eqnarray}
where $s_j, \ t_j\in{\mathbb C}, \ j=1,2,3$. We require that the
eigenvectors and eigenforms are dual to each other
\cite{mf89,cc05,mf81}, namely, $f_ju_k = f_j^* u_k^* = \delta_{jk},
\ f_ju_k^* = f_j^*u_k = 0$, implying that $s_1=\frac{1}{4t_1}$,
$s_2=\frac{1}{4t_2}$, $s_3=\frac{1}{2t_3}$. The constants $t_j$ will
be fixed later to simplify some commutation relationships. Finally,
the eigenvectors and eigenforms satisfy the unit matrix
decomposition
\begin{equation}
\mathbf{1}= \sum_{k=1}^3 \left(u_k\otimes f_k +
                    u_k^* \otimes  f_k^*\right) \quad \Rightarrow  \quad
                    \pmb{\Lambda} = \sum_{k=1}^3 \lambda_k \left(u_k\otimes f_k
                    - u_k^* \otimes  f_k^*\right) \label{sdl}
\end{equation}
$\otimes$ denoting tensor product. The $\pmb{\Lambda}$-expression in
(\ref{sdl}) allows to decompose the Heisenberg trajectories as three
oscillating modes of frequencies $\omega_j$ \cite{mf89,cc05}.
Moreover, it will characterize as well the algebraic structure of
the Hamiltonian.

\section{Algebraic structure of $H$}

We can define now three pairs of ladder operators  of $H$, $L_k=
f_k^*\eta$, $L_k^\dag = f_k \eta$, $k=1,2,3$, which obey the
following commutation relations with $H$:
\begin{eqnarray}\label{[H,L's]}
  &&  [H,L_k] = -i f_k^* [iH,\eta] =
    -\omega_k L_k, \qquad
[H,L_k^\dag] = \omega_k L_k^\dag.
\end{eqnarray}
An explicit calculation leads to:
\begin{eqnarray}
\nonumber
    & L_1 = t_1^*\left[\sqrt{b^2+v}(X-iY)+i(P_x-iP_y)\right], \\
    & \hskip-0.9cm L_2 =
    t_2^*\left[-\sqrt{b^2+v}(X-iY)+i(P_x-iP_y)\right],
    \label{ladops} \quad
    L_3 =  t_3^*\left(-i\sqrt{-2v}Z+P_z \right)
    \label{ladderexplicit}.
\end{eqnarray}
By evaluating next the commutators between $L_i, L_j^\dagger$, the
following non-null results are obtained:
\begin{eqnarray}
    \nonumber
    & [L_1,L_1^\dag] = 2|t_1|^2 (\omega_1 - \omega_2)= 1,\\
    &
    [L_2,L_2^\dag] = - 2|t_2|^2 (\omega_1 - \omega_2)= -
    1, \quad [L_3,L_3^\dag] = 2|t_3|^2 \omega_3 = 1
    \label{Lcommutators},
\end{eqnarray}
where we have finally chosen $t_i\in{\mathbb R}^+$ such that $t_1 =
t_2 = 1/\sqrt{2(\omega_1 - \omega_2)}$, $t_3 = 1/\sqrt{2\omega_3}$
to simplify at maximum equation (\ref{Lcommutators}). On the other
hand, $ [L_i^\dag,L_j^\dag]=[L_i,L_j]=0$, $i,j=1,2,3$.

Now $H$ is factorized in terms of $L_k, \ L_k^\dagger$ as follows
\cite{mf89,cc05}:
\begin{eqnarray}
& H = \omega_1 L_1^\dagger L_1 - \omega_2 L_2 L_2^\dagger + \omega_3
L_3^\dagger L_3 + (\omega_1 - \omega_2 + \omega_3)/2. \label{h3o}
\end{eqnarray}
Moreover, equations (\ref{Lcommutators},\ref{h3o}) allow to identify
three independent oscillator modes for $H$, each one characterized
by its number $N_k$, annihilation $B_k$ and creation $B_k^\dagger$
operator, in the way:
\begin{eqnarray}
& N_k = B_k^\dagger B_k, \quad k=1,2,3, \label{npbbd}  \\
& B_1 = L_1, \quad B_2 = L_2^\dagger,  \quad B_3 = L_3,
\label{opsann} \quad B_1^\dagger = L_1^\dagger, \quad B_2^\dagger =
L_2 , \quad B_3^\dagger = L_3^\dagger.
\end{eqnarray}
They obey the standard commutation relations:
\begin{eqnarray}
& [N_k,B_k] = - B_k, \quad [N_k, B_k^\dagger] =  B_k^\dagger, \quad
[B_j,B_k^\dagger] = \delta_{jk}, \quad j,k=1,2,3.
\label{Bcommutators}
\end{eqnarray}
Hence, one can construct a basis $\{ \vert n_1,n_2,n_3 \rangle, \
n_j=0,1,2,\dots, j=1,2,3\}$ of common eigenstates of $N_1, \ N_2, \
N_3$,
\begin{eqnarray}
& N_j \vert n_1,n_2,n_3 \rangle = n_j \vert n_1,n_2,n_3 \rangle,
\quad j=1,2,3,
\end{eqnarray}
departing from an {\it extremal state} $\vert 0,0,0 \rangle$ which
is annihilated by $B_1, \ B_2, \ B_3$:
\begin{eqnarray}
& B_j \vert 0,0,0 \rangle = 0, \quad j = 1,2,3.
\end{eqnarray}
If we assume that $\vert 0,0,0 \rangle$ is normalized, it turns out
that \cite{ve07}:
\begin{eqnarray}
& \vert n_1,n_2,n_3 \rangle = (n_1! \, n_2! \, n_3!)^{-1/2}
B_1^\dagger{}^{n_1} \, B_2^\dagger{}^{n_2} \, B_3^\dagger{}^{n_3}
\vert 0,0,0 \rangle .
\end{eqnarray}
Moreover, $B_j, \ B_j^\dagger, \ j = 1,2,3$ act onto $\vert
n_1,n_2,n_3 \rangle$ in a standard way:
\begin{eqnarray}
& \hskip-0.8cm  B_1 \vert n_1,n_2,n_3 \rangle =  \sqrt{n_1} \, \vert
n_1 \! - \! 1,n_2,n_3 \rangle, \nonumber \ B_1^\dagger \, \vert
n_1,n_2,n_3 \rangle  =  \sqrt{n_1 \! + 1} \, \vert n_1 \! +
1,n_2,n_3 \rangle,
\end{eqnarray}
and similar expressions for the action of $B_2, \ B_2^\dagger, \
B_3, \ B_3^\dagger$. Notice that $\vert n_1,n_2,n_3 \rangle$ is
eigenstate of the Penning trap Hamiltonian with eigenvalue
$E_{n_1,n_2,n_3} = \omega_1 (n_1 + 1/2) - \omega_2 (n_2 + 1/2)  +
\omega_3 (n_3 + 1/2) \equiv E(n_1,n_2,n_3)$. In particular, the
extremal state $\vert 0,0,0 \rangle$ has eigenvalue $E_{0,0,0} =
(\omega_1 - \omega_2 + \omega_3)/2$, i.e., it is neither a ground
nor a top state since its energy is ``in the middle'' of the
spectrum of $H$. Following \cite{fhr07}, it is seen that there is an
{\it intrinsic} algebraic structure for our system, which is
characterized by a linear relationship between the Penning trap
Hamiltonian $H$ and the three number operators $N_k$:
\begin{eqnarray}
& H = E(N_1,N_2,N_3) = \omega_1 N_1  - \omega_2 N_2 +  \omega_3 N_3
+ E_{0,0,0} . \label{intrinsicE(N)}
\end{eqnarray}
As it happens for one-dimensional systems, in our three-dimensional
example the detailed structure is contained in the operator relation
(\ref{intrinsicE(N)}), which is responsible of the specific spectrum
and, consequently, of  the lack of a ground or a top proper energy.
On the other hand, the global structure comes from the very
existence of the three independent oscillator modes for $H$, each
one characterized by the standard generators $\{N_j, \ B_j, \
B_j^\dagger\}, \ j = 1,2,3$. This global behavior allows to identify
in a natural way the extremal state $\vert 0,0,0\rangle$ which,
although is neither a ground nor a top energy eigenstate, plays the
same role as the ground state for the one-dimensional harmonic
oscillator.

\section{Extremal state wave function}

The existence of the extremal state $|0,0,0\rangle$ is guaranteed by
a theorem which is proven elsewhere \cite{mf89}. It ensures that, if
the operators
\begin{eqnarray} \label{desarrollo B en alfa y beta}
B_j=i~\vec{P}\cdot\vec{\alpha}_j +\vec{R}\cdot\vec{\beta}_j,
        \quad
B_j^\dagger=-i~\vec{\alpha}_j^\dagger \cdot \vec{P} +
\vec{\beta}_j^\dagger \cdot
        \vec{R}, \quad  j=1,2,3,
\end{eqnarray}
obey the commutation relations (\ref{Bcommutators}), then the system
of partial differential equations $\langle \vec{r}| B_j
|0,0,0\rangle=0, \ j=1,2,3,$ for the extremal state wave function
$\phi_{\mathbf 0}(\vec{r}) \equiv \langle \vec{r}|0,0,0\rangle$ has a
square integrable solution given by
\begin{eqnarray} \label{solucion sistema de ec}
\phi_{\mathbf 0}(\vec{r}) = c \exp\left(-\frac{1}{2}a_{ij}r_i
r_j\right) = c
    \exp\left(-\frac{1}{2}\vec{r}\,^{\rm T}\mathbf{a}\vec{r}\right),
\end{eqnarray}
where $\mathbf{a}=(a_{ij})$ is a complex symmetric matrix satisfying
\begin{eqnarray} \label{alfa a igual a beta}
    \mathbf{a}\vec{\alpha}_j=\vec{\beta}_j, \quad
    j=1,2,3.
\end{eqnarray}
According to (\ref{desarrollo B en alfa y beta}), through equations
(\ref{ladderexplicit},\ref{opsann}) we identify the vectors
\begin{eqnarray} \nonumber
\vec{\alpha}_1 &=&\frac{1}{2(b^2+v)^{1/4}}\left(
                 1, \ -i,  \ 0
             \right)^{\rm T}, \qquad \vec{\beta}_1 =
             (b^2+v)^{1/2}\vec{\alpha}_1,
             \\
    \vec{\alpha}_2 &=&-\frac{1}{2(b^2+v)^{1/4}}\left(
                 1, \ i, \ 0
             \right)^{\rm T}, \qquad \vec{\beta}_2 =
             (b^2+v)^{1/2}\vec{\alpha}_2, \label{alphaexpression} \\ \nonumber
             \vec{\alpha}_3 &=&-\frac{i}{\sqrt{2}(-2v)^{1/4}}\left(
               0, \ 0, \ 1
             \right)^{\rm T} , \qquad \vec{\beta}_3 = (-2v)^{1/2}\vec{\alpha}_3.
\end{eqnarray}
Thus, $\mathbf{a}= {\rm
diag}\left[\sqrt{b^2+v},\sqrt{b^2+v},\sqrt{-2v}\right]$, and from
(\ref{solucion sistema de ec}) we finally get the extremal state
wave function we were looking for:
\begin{eqnarray}\label{edo extremal caso I}
    \phi_{\mathbf 0}(\vec{r})=c\exp\left({-\frac{\sqrt{b^2+v}}{2}(x^2+y^2)-
    \sqrt{\frac{-v}{2}}~z^2}\right).
\end{eqnarray}

\section{Penning trap coherent states}

Once the Penning trap Hamiltonian has been expressed appropriately
in terms of annihilation and creation operators, we can develop a
similar treatment as for the harmonic oscillator to build up the
corresponding coherent states.

\subsection{Annihilation operator coherent states}

In the first place let us look for the annihilation operator
coherent states (AOCS) as common eigenstates of $B_1, \ B_2, \ B_3$:
\begin{eqnarray}
 & B_j \vert
z_1,z_2,z_3 \rangle = z_j \vert z_1,z_2,z_3 \rangle, \quad j=1,2,3.
\label{cseB}
\end{eqnarray}
Following a standard procedure, let us expand them in the basis $\{
\vert n_1,n_2,n_3 \rangle \}$:
\begin{eqnarray}
&& \vert z_1,z_2,z_3 \rangle = \sum_{n_1,n_2,n_3 = 0}^\infty
c_{n_1,n_2,n_3} \vert n_1,n_2,n_3 \rangle.
\end{eqnarray}
By asking that (\ref{cseB}) is satisfied, three recurrence
relationship for $c_{n_1,n_2,n_3}$ will be obtained, which in turn
lead to the following expressions:
\begin{eqnarray}
& \hskip-0.8cm c_{n_1,n_2,n_3} =(n_1!)^{-1/2} z_1^{n_1}c_{0,n_2,n_3}
= (n_2!)^{-1/2} z_2^{n_2}c_{n_1,0,n_3} = (n_3!)^{-1/2} z_3^{n_3}
c_{n_1,n_2,0}.
\end{eqnarray}
Hence, it is straightforward to show that
\begin{eqnarray}
c_{n_1,n_2,n_3} = (n_1!\ n_2! \ n_3!)^{-1/2}
z_1^{n_1}z_2^{n_2}z_3^{n_3}c_{0,0,0},
\end{eqnarray}
where $c_{0,0,0}$ is to be found from the normalization condition.
Thus, up to a global phase factor, the normalized AOCS become
finally:
\begin{eqnarray}
&& \hskip-0.8cm \vert z_1,z_2,z_3 \rangle = e^{-\frac{\vert
z_1\vert^2 + \vert z_2\vert^2 + \vert z_3\vert^2}2}\sum_{n_1,n_2,n_3
= 0}^\infty (n_1!\ n_2! \ n_3!)^{-1/2}
z_1^{n_1}z_2^{n_2}z_3^{n_3}\vert n_1,n_2,n_3 \rangle. \label{cspta}
\end{eqnarray}

\subsection{Displacement operator coherent states}

According to equation (\ref{gsdisplaced}), for the $j$-th mode of
the Penning trap Hamiltonian we have to take into account the
corresponding displacement operator $D_j(z_j)=\exp(z_jB^\dagger_j -
z_j^* B_j)$. By using the BCH formula it turns out that:
\begin{eqnarray}
&   D_j(z_j)=e^{-\frac{\vert z_j\vert^2}2} e^{z_j B_j^\dagger}
e^{-z_j^*B_j}, \quad j=1,2,3.
\end{eqnarray}
Now, the global displacement operator is given by:
\begin{eqnarray}
    D(\mathbf{z}) \equiv D(z_1,z_2,z_3)= D_1(z_1)D_2(z_2)D_3(z_3),
\end{eqnarray}
where $\mathbf{z}$ denotes the complex variables $z_1, z_2, z_3$
associated to the three modes. By employing now the second
definition, we get the displacement operator coherent states (DOCS)
$|{\mathbf{z}}\rangle$ from applying $D(\mathbf{z})$ to the extremal
state $\vert 0,0,0 \rangle$:
\begin{eqnarray}
\hskip-0.5cm    |{\mathbf{z}}\rangle& = & D(\mathbf{z})|0,0,0\rangle
=  e^{-\frac{\vert z_1\vert^2 + \vert z_2\vert^2 +
    \vert z_3\vert^2}{2}}\sum_{n_1,n_2,n_3 = 0}^{\infty}
    \frac{z_1^{n_1}z_2^{n_2}z_3^{n_3}|n_1,n_2,n_3\rangle}{\sqrt{n_1!~n_2!~n_3!}}.
    \label{z de D en z=0}
\end{eqnarray}
By comparing (\ref{cspta}) and (\ref{z de D en z=0})) we realize
that the DOCS and the AOCS are the same. Moreover, since
$[z_jB_j^\dagger - z_j^* B_j,z_k B_k^\dagger-z_k^* B_k]=0$,
$j,k=1,2,3$, we get
\begin{eqnarray}\nonumber
    & \hskip-2cm D(\mathbf{z}) \! = \! \exp(z_1 B_1^\dagger+z_2 B_2^\dagger+ z_3 B_3^\dagger
    -z_1^*B_1 - z_2^*B_2 - z_3^*B_3) \! = \! \exp[i(\vec\Sigma \! \cdot \! \vec{R}
    - \vec{\Gamma}\! \cdot \! \vec{P})] \\
    &\hskip-1.2cm = C(\mathbf{z})
    F(\vec{R}) \exp(-i\vec{\Gamma}\cdot\vec{P})
    =[C(\mathbf{z})]^{-1}
    \exp(-i\vec{\Gamma}\cdot\vec{P})
    F(\vec{R}), \label{D expandido}
\end{eqnarray}
where we have used the BCH formula and equation (\ref{desarrollo B
en alfa y beta}) to identify
\begin{eqnarray}
 && \hskip-1cm  \vec{\Gamma} = \left(
                                                                  \begin{array}{l}
    (b^2+v)^{-\frac{1}{4}} \ {\rm{Re}}[z_1-z_2] \\
    -(b^2+v)^{-\frac{1}{4}}\ {\rm{Im}}[z_1+z_2] \\
    -(-v/2)^{-\frac{1}{4}}~{\rm{Im}}[z_3]
\end{array}
                                                               \right),
                                                               \quad
    \vec{\Sigma} = \left(
                                                                 \begin{array}{l}
    (b^2+v)^{\frac{1}{4}}\ {\rm{Im}}[z_1-z_2] \\
    (b^2+v)^{\frac{1}{4}}\ {\rm{Re}}[z_1+z_2] \\
    (-8v)^{\frac{1}{4}}{\rm{Re}}[z_3]
\end{array}
                                                               \right),
\label{vec Gama y Sigma} \\
&& \hskip-1cm  C(\mathbf{z}) = e^{-i\vec{\Gamma}\cdot\vec{\Sigma/2}}
    = \exp\{i({\rm{Re}}[z_1]{\rm{Im}}[z_2]
    + {\rm{Re}}[z_2]{\rm{Im}}[z_1] + {\rm{Re}}[z_3]{\rm{Im}}[z_3])\}, \nonumber \\
\label{D definicion F(x)}
   && \hskip-1cm  F(\vec{R}) \! = \! e^{i \vec{\Sigma}  \cdot
   \vec{R}} \!
    = \exp\{i(b^2 \! + v)^{\frac{1}{4}}
            ({\rm{Im}}[z_1 \! - \! z_2]X
                    \! + \! {\rm{Re}}[z_1 \! + \!z_2]Y)
            \! + \! i (-8v)^{\frac{1}{4}}{\rm{Re}}[z_3]Z\}.
            \nonumber
\end{eqnarray}
Since the operator $e^{-i\vec{P}\cdot\vec{\Gamma}}$, $\Gamma\!_i \in
\mathbb{R}$, performs a coordinate displacement in the way $\langle
\vec{r}|e^{-i\vec{P}\cdot\vec{\Gamma}}= \langle
\vec{r}-\vec{\Gamma}|$, we finally get:
\begin{eqnarray}\nonumber
    \hskip-1.5cm \phi_\mathbf{z}(\vec{r})& \equiv &\langle \vec{r}|\mathbf{z}\rangle =
    \langle \vec{r}|D(\mathbf{z})\vert 0,0,0\rangle
    = C(\mathbf{z})F(\vec{r})
    \langle \vec{r}|
    e^{-i\vec{P}\cdot\vec{\Gamma}}
    |0,0,0\rangle, \\
\label{finalcswf}
 \hskip-1.5cm & = &
     C(\mathbf{z})F(\vec{r})
    \phi_\mathbf{0}\left(x-\frac{{\rm{Re}}[z_1 - z_2]}{(b^2+v)^{\frac{1}{4}}},
    y+\frac{{\rm{Im}}[z_1 + z_2]}{(b^2+v)^{\frac{1}{4}}},
    z+\left(\frac{-2}{v}\right)^{\frac{1}{4}}\rm{Im}[z_3]\right),
\end{eqnarray}
with $\phi_\mathbf{0}(\vec{r})$ given by (\ref{edo extremal caso
I}).


\section{Mean values of physical quantities}

Let us evaluate next the mean values $\langle R_j\rangle_\mathbf{z}
\equiv \langle \mathbf{z}|R_j|\mathbf{z}\rangle$, $\langle
P_j\rangle_\mathbf{z}\equiv \langle
\mathbf{z}|P_j|\mathbf{z}\rangle$, $j=1,2,3$, and the corresponding
mean square deviations in a given CS $|\mathbf{z}\rangle$. To do
that, we analyze first how the operators $R_j, \ R_j^2, \ P_j, \
P_j^2$ are transformed under $D(\mathbf{z})$. By using equations
(\ref{D expandido}) and (\ref{vec Gama y Sigma}) it is
straightforward to show that:
\begin{eqnarray}
\hskip-0.8cm && D^\dagger(\mathbf{z}) R_j^n D(\mathbf{z}) \! = \!
(R_j + \Gamma\!_j)^n, \quad D^\dagger(\mathbf{z}) P_j^n
D(\mathbf{z}) \! = \! (P_j + \Sigma_j)^n, \ n \! = \! 1,2,\dots
\end{eqnarray}
Therefore:
\begin{eqnarray}
   & \hskip-0.9cm   \langle R_j\rangle_\mathbf{z} =
    \langle R_j \rangle_\mathbf{0} + \Gamma\!_j,  \label{evxjz} \
    \langle R_j^2\rangle_\mathbf{z} =  \langle R_j^2\rangle_\mathbf{0} + 2\Gamma\!_j\langle
    R_j\rangle_\mathbf{0} + {\Gamma\!_j}^2, \label{valor esperado X^2_j}
    \
    (\Delta R_j)^2_\mathbf{z}
    =(\Delta R_j)_\mathbf{0}^2 \label{Delta X_z = Delta X_0}, \\
    & \hskip-1cm  \langle P_j \rangle_\mathbf{z}= \langle
    P_j\rangle_\mathbf{0} + \Sigma_j,
    \label{valor esperado P_j} \
    \langle P_j^2\rangle_\mathbf{z}
    = \langle P_j^2\rangle_\mathbf{0} + 2\Sigma_j\langle
    P_j\rangle_\mathbf{0} + {\Sigma_j}^2, \label{valor esperado P^2_j}
    \
    (\Delta P_j)_\mathbf{z}^2 = (\Delta P_j)_\mathbf{0}^2.
    \label{Delta P_z = Delta P_0}
\end{eqnarray}
Notice that the mean square deviations of $R_j$ and $P_j$ are
independent of $z_1$, $z_2$, $z_3$ but depend on $\langle
R_j\rangle_\mathbf{0}$, $\langle P_j\rangle_\mathbf{0}$, $\langle
R_j^2\rangle_\mathbf{0}$, $\langle P_j^2\rangle_\mathbf{0},$
$j=1,2,3$, which need to be evaluated. The first six quantities can
be obtained from the homogeneous equations $\langle B_k
\rangle_\mathbf{0} = i(\vec{\alpha}_k)_j\langle P_j
\rangle_\mathbf{0} +
    (\vec{\beta}_k)_j\langle R_j\rangle_\mathbf{0} = 0$, $\langle B_k^\dagger
\rangle_\mathbf{0} = -i(\vec{\alpha}_k^*)_j\langle P_j
\rangle_\mathbf{0} + (\vec{\beta}_k^*)_j\langle
    R_j\rangle_\mathbf{0} = 0, \ k=1,2,3$ (see (\ref{desarrollo B en
alfa y beta}) and use that $B_k|0,0,0\rangle = \langle 0,0,0 \vert
B_k^\dagger = 0$). By using (\ref{alphaexpression}), the system to
be solved becomes:
\begin{eqnarray}
 &   -i\sqrt{-2v}\langle Z\rangle_\mathbf{0} + \langle P_z\rangle_\mathbf{0} = 0, \nonumber \\
    \nonumber
 &   \sqrt{b^2+v}\left(\langle X\rangle_\mathbf{0} - i\langle Y\rangle_\mathbf{0} \right)
            +i\left(\langle P_x\rangle_\mathbf{0} - i\langle P_y\rangle_\mathbf{0} \right) = 0, \\
    \nonumber
 &   -\sqrt{b^2+v}\left(\langle X\rangle_\mathbf{0} + i\langle Y\rangle_\mathbf{0}\right)
            -i\left(\langle P_x\rangle_\mathbf{0} + i\langle P_y\rangle_\mathbf{0}\right) = 0,
\end{eqnarray}
and the complex conjugate equations. Its solution is given by
\begin{eqnarray}
    \langle R_j \rangle_\mathbf{0} = \langle P_j
    \rangle_\mathbf{0} = 0, \quad j = 1,2,3.
\end{eqnarray}
In order to obtain $\langle R_j^2 \rangle_\mathbf{0}$, $\langle
P_j^2 \rangle_\mathbf{0}$, we calculate the mean values for the
several products of pairs involving $B_j, \ B_k^\dagger$. From these
thirty six equations just twenty one are linearly independent:
$\langle B_j B_k \rangle_\mathbf{0} = 0, \ j=1,2,3,k\leq j$ (six
equations); $\langle B_j^\dagger B_k ^\dagger\rangle_\mathbf{0} = 0,
\ j=1,2,3,k\leq j$ (six equations); $\langle B_k^\dagger
B_j\rangle_\mathbf{0} = 0, \ j,k=1,2,3,$ (nine equations). By
solving this linear system, the non-null mean values of the twenty
one independent products of $R_i$ and $P_j$ are now:
\begin{eqnarray}\nonumber
    && \langle X^2\rangle_\mathbf{0} =
    \langle Y^2 \rangle_\mathbf{0} = [4(b^2+v)]^{-\frac{1}{2}},
        \qquad
    \langle Z^2 \rangle_\mathbf{0}=(-8v)^{-\frac{1}{2}}, \\ \nonumber
    && \langle P_x^2\rangle_\mathbf{0}=
    \langle P_y^2 \rangle_\mathbf{0} = [(b^2+v)/4]^{\frac{1}{2}} ,
         \qquad
    \langle P_z^2 \rangle_\mathbf{0} = (-v/2)^{\frac{1}{2}}, \\
    \nonumber
    && \langle XP_x\rangle_\mathbf{0} =
    \langle YP_y\rangle_\mathbf{0} =
    \langle ZP_z\rangle_\mathbf{0} = i/2.
\end{eqnarray}
The previous formulas imply that equations (\ref{Delta X_z = Delta
X_0},\ref{Delta P_z = Delta P_0}) become
\begin{eqnarray}\nonumber
    (\Delta X)_\mathbf{z}^2=(\Delta Y)_\mathbf{z}^2&=&[4(b^2+v)]^{-\frac{1}{2}},
    \quad
    (\Delta Z)_\mathbf{z}^2 = (-8v)^{-\frac{1}{2}},
    \\ \nonumber
    (\Delta P_x)_\mathbf{z}^2=(\Delta P_y)_\mathbf{z}^2&=&
        [(b^2+v)/4]^{\frac{1}{2}},
    \quad
    (\Delta P_z)_\mathbf{z}^2=(-v/2)^{\frac{1}{2}},
\end{eqnarray}
and therefore
\begin{eqnarray}\nonumber
     (\Delta X)_\mathbf{z}(\Delta P_x)_\mathbf{z}=
     (\Delta Y)_\mathbf{z}(\Delta P_y)_\mathbf{z}=
     (\Delta Z)_\mathbf{z}(\Delta P_z)_\mathbf{z}= 1/2.
\end{eqnarray}
This means that our CS have minimum Heisenberg uncertainty
relations.

Finally, by using equations (\ref{npbbd},\ref{intrinsicE(N)}) we
calculate the mean value of the Hamiltonian $H$ in a given CS
$|\mathbf{z}\rangle$:
\begin{eqnarray}\label{meanvalueH}
    \langle H \rangle_\mathbf{z}=
    \omega_1 |z_1|^2  - \omega_2 |z_2|^2 + \omega_3 |z_3|^2 + E_{0,0,0} .
\end{eqnarray}
A similar calculation for $\langle H^2 \rangle_\mathbf{z}$ can be
done, leading to:
\begin{equation}
    (\Delta
    H)^2_\mathbf{z}=\left(b+\sqrt{b^2+v}\,\right)^2|z_1|^2 +
    \left(b - \sqrt{b^2+v}\,\right)^2|z_2|^2 - 2v|z_3|^2.
\end{equation}
Once again, the fact that $H$ is not positive definite is clearly
reflected in (\ref{meanvalueH}).

Along this work we have assumed that $b = - \frac{eB}{2mc}>0$. For
$b<0$, small differences concerning the identification of the
appropriate annihilation and creation operators arise. However, the
extremal state and CS wave functions $\phi_\mathbf{0}(\vec{r})$,
$\phi_\mathbf{z}(\vec{r})$ as well as the corresponding mean values,
will coincide with those previously calculated. In particular, the
Heisenberg uncertainty relation will achieve once again its minimum
value \cite{ve07}.

\section{Concluding remarks}\label{C:conclusiones}

In this paper it was introduced a technique to find the CS for a
charged particle in an ideal Penning trap. We have shown that the
coherent states, calculated through both definitions given by
equations (\ref{eigena},\ref{gsdisplaced}), are the same. We
introduced also a prescription to obtain the mean values of several
physical observables in a given coherent state. We have found,
finally, that the Penning trap coherent states (derived
algebraically) obey also the third CS definition, i.e., they satisfy
the minimum Heisenberg uncertainty relation.

Let us remark that the method presented here is quite general, and
it could be applied to other systems characterized by quadratic
Hamiltonians. In order to implement systematically this treatment,
we have to identify first the stability regions where the
non-degenerate eigenvalues of $\pmb{\Lambda}$ become purely
imaginary, which ensures that the Heisenberg and classical
trajectories are trapped. In the trap regime the Hamiltonian is
decomposed in terms of independent harmonic oscillators, and thus
our procedure can be straightforwardly applied. Note that
generalizations of this kind have been elaborated elsewhere (see
e.g. \cite{dm89}). However, in our method it is direct to identify
the global sign accompanying each individual oscillator involved in
the Hamiltonian decomposition. As we saw in our Penning trap
example, those signs determine the existence or not of a ground
state for the system, a fact which is not well known in the
literature. Moreover, they become fundamental for the determination
of the intrinsic algebraic structure of the involved Hamiltonian
(compare equation (\ref{intrinsicE(N)})). Observe that some of these
properties were found previously for operators imitating the
Hamiltonian in non-inertial reference frames \cite{mf89,cc05}. By
means of this example we have shown that such a property arises as
well for Hamiltonians in inertial frames of reference.

\section*{Acknowledgments}{The authors acknowledge the support of Conacyt,
project No. 49253-F. MV acknowledges to Conacyt a MSc grant as well
as the support of Cinvestav.}


\begin{thebibliography}{00}

\bibitem{ks85}
JR Klauder, BS Skagerstam~Eds, \emph{Coherent states. Applications
in physics and mathematical physics}, World Scientific, Singapore
(1985)

\bibitem{pe86}
A~Perelomov, \emph{Generalized coherent states and their
applications}, Springer-Verlag, Heidelberg (1986); WM Zhang, DH
Feng, R Gilmore, Rev Mod Phys {\bf 62} (1990) 867

\bibitem{gaa00}
JP~Gazeau, ST~Ali, JP~Antoine, \emph{Coherent states, wavelets and
their generalizations}, Springer-Verlag, New York (2000); VV
Dodonov, J Opt B \textbf{4} (2002) R1

\bibitem{rr00}
B Roy, P Roy, J Op B {\bf 2} (2000) 65; C Quesne, Ann Phys {\bf 293}
(2001) 147; J Recamier, PG de Le\'on, R J\'auregui, A Frank, O
Casta\~nos, Int J Quant Chem {\bf 89} (2002) 494

\bibitem{mf89}
B Mielnik, DJ~Fern\'andez, J Math Phys \textbf{30} (1989) 537; DJ
Fern\'andez, Acta Phys Polon \textbf{B21} (1990) 589

\bibitem{cc05}
S~Cruz y~Cruz, PhD Thesis, Cinvestav (2005); S Cruz y Cruz, B
Mielnik, Phys Lett A {\bf 352} (2006) 36

\bibitem{bg86}
LS Brown, G Gabrielse, Rev Mod Phys {\bf 58} (1986) 233

\bibitem{fe92}
DJ Fern\'andez, LM Nieto, Phys Lett A {\bf 157} (1991) 315; DJ
Fern\'andez, Nuovo Cim {\bf 107B} (1992) 885; DJ Fern\'andez, N
Bret\'on, Europhys Lett {\bf 21} (1993) 147

\bibitem{fm94}
DJ Fern\'andez, B Mielnik, J Math Phys {\bf 35} (1994) 2083

\bibitem{ma97} OV Manko, Phys Lett A {\bf 228} (1997) 29;
O Casta\~nos, S Hacyan, R L\'opez-Pe\~na, VI
Manko, J Phys A {\bf 31} (1998) 1227

\bibitem{gl63a}
RJ Glauber, Phys Rev Lett \textbf{10} (1963) 84; RJ Glauber, Phys
Rev \textbf{131} (1963) 2766

\bibitem{bbgk71}
V Bargmann,  P~Butera, L~Girardello, JR~Klauder, Rep Math Phys
\textbf{2} (1971) 221

\bibitem{mf81}
JV~Moloney, FHM~Faisal, J Phys B \textbf{14} (1981) 3603

\bibitem{ve07}
MP Vel\'azquez Quesada, MSc Thesis, Cinvestav (2007)

\bibitem{fhr07}
DJ Fern\'andez, V Hussin, O Rosas-Ortiz, J Phys A {\bf 40} (2007)
6491

\bibitem{dm89}
VV Dodonov, VI Manko, \emph{Invariants and the evolution of
nonstationary quantum systems}, MA Markov Ed, Nova Science, New York
(1989)

\end{thebibliography}
\end{document}